\documentclass[%
reprint,
superscriptaddress,
nofootinbib,
amsmath,amssymb,
aps,
prab,
]{revtex4-2}

\usepackage{graphicx}
\usepackage{dcolumn}
\usepackage{bm}
\usepackage{siunitx}
\usepackage{physics}
\usepackage{soul}

\begin{document}


\title{Design of a resonant slow extraction from low emittance electron booster rings using transverse resonance island buckets}

\author{E.C. Cortés García}
\email{edgar.cristopher.cortes.garcia@desy.de}
\affiliation{Deutsches Elektronen-Synchrotron DESY, Notkestr. 85, 22607 Hamburg, Germany }

\author{I.V. Agapov}%
\affiliation{%
Deutsches Elektronen-Synchrotron DESY, Notkestr. 85, 22607 Hamburg, Germany 
}%
\author{W. Hillert}
\affiliation{Universit\"at Hamburg, Fachbereich Physik, Institut für Experimentalphysik, Luruper Chaussee 149, 22761 Hamburg}

\date{\today}

\begin{abstract}
In this contribution we present the design of a resonant slow extraction based on the radio frequency knock-out (RF-KO) scheme, where we make use of the transverse resonance islands bucket (TRIB) optics.\ 
The generation of the TRIB optics is presented in two example lattices, that are considered for the potential upgrade of the current booster at DESY.\
The slow extraction would potentially supply high energy electron beams to the HEP beamline users.\ 
Simulations show an extraction efficiency in excess of 90\% with a septum blade thickness of \SI{100}{\micro\meter}.
\end{abstract}

\maketitle


\section{Introduction}

The resonant slow extraction technique from synchrotrons and storage rings is widely employed to deliver a variety of beams, typically for fixed-target experiments.\
This method involves bringing the machine's tune close to a resonance point, deliberately reducing the dynamic aperture.\
As a result, particles are driven into unstable motion, facilitating their easy ejection into an extraction channel~\cite{strolin1966,Barton:1971,Steinbach:1993,Hardt:1981}.\ 
In recent implementations of the method the third order resonance has been established as the standard working point for synchrotron facilities.\ 
The resonance drivers are sextupoles, which are already commonly found for chromaticity correction, or dedicated sextupole families that are installed at dispersion-free regions for this purpose.\

Furthermore, the rate at which particles are extracted depends on how the beam is brought to meet the resonance condition.\
In recent years, considerable improvements have been made in the capability to control the extracted particle rate, largely due to applications in ion therapy~\cite{medsyncNoda, medsyncPullia}, where requirements are more stringent.\
Therefore, an extensive set of tools has been made available for the radio frequency knock-out (RF-KO) scheme~\cite{TOMIZAWA1993} to precisely control the rate of slowly extracted ion particles.\
This is achieved by shaping the excitation waveform used for extraction and active control of the excitation strength with feedback and feedforward loops~\cite{NodaRFwithAFM,NodaDualFM,NodaExtDiffRegion,FurukawaSyncOszi,NakanishiMultiModeSim,YamaguchiMultiMode,SchoemersFeedbackSystem, FURUKAWA2004}.

Slow resonant extraction is widely implemented in facilities producing hadron beams but is rarely available at electron synchrotrons\footnote{In fact, the only facility offering a resonant slowly extracted electron beam is ELSA at Bonn \cite{GentnerPhD}.}.\
For this reason, the inherent properties of low-emittance electron machine lattices have been seldom discussed in the context of resonant slow extraction.\

In low-emittance lattices, the emittance is optimized by suppressing dispersion and minimizing the average $\mathcal{H}$ function \cite{WolskiLowEmittanceRings, SYLeeBook}.\
Achieving this typically involves reducing the cell length while increasing the phase advance, resulting in high quadrupole gradients.\
A notable consequence of this setup is a high horizontal tune and an even higher natural chromaticity.\

\ {Consequently, strong sextupoles are installed to correct the natural chromaticity and mitigate collective and single-particle instabilities.}\ These sextupoles significantly amplify nonlinear features in transverse dynamics.\
As a result, higher-order multiple components cannot be disregarded when describing the system's dynamics.

In this contribution, we explore the possibility of slowly extracting a high-quality electron beam from a low-emittance booster ring.\ 
We propose to set up the slow extraction optics such that stable transverse resonance island buckets (TRIBs) are generated and show that by slowly pushing particles to the external islands a reasonable extraction efficiency of more than $92\%$ with a \SI{100}{\micro\meter} thick effective septum blade can be achieved.\
\ {This uncommon optics setup is motivated by the fact that in the low emittance lattices we studied the usual resonant slow extraction optics was not attainable without turning off the chromatic sextupoles}.\ 
\ {Due to the expected equilibrium momentum spread, the correction of chromatic tune spread with strong sextupoles is necessary}.\
\ {This might be a common characteristic of modern low emittance booster rings and very large machines.}

Further, optics that don't lead to generation of TRIBs appear to only constraint the dynamic aperture. A particle that leaves the dynamic aperture, simply follows chaotic motion
and gets lost in a very short time. Unlike the conventional extraction, there is no separatrix arm, where the particles’ orbits converge to. While the design of resonant slow extraction has been extensively discussed in previous studies (e.g. \cite{benediktPHD, pulliaPHD, PIMMS-Study, CASBryant:2018}), \ {slow extraction with TRIBs optics has not yet been considered.}\ 
When performing the chromaticity correction, sextupoles induce an amplitude-dependent tune shift (ADTS).\
The consequences of including this term in the dynamics, while simultaneously operating near a third-order resonance, have been investigated along two parallel fronts.\
First, within the resonant slow extraction community, higher-order multipoles have been introduced to minimize the activation of components in the extraction channel \cite{FraserPRAB:2019, NagaslaevPRAB:2019}.\
Second, the study of stable transverse resonance island buckets (TRIBs) \cite{PRL2002_CappiGiovannozzi, Borburgh_2016}, particularly those generated near the third-order resonance in electron storage rings \cite{Ries:IPAC2015-MOPWA021, CornellTRIBS_PRAB23}, has been pursued.

There are numerous ways to implement slow extraction, each corresponding to various methods for bringing a particle into resonance.\
An extensive discussion of these possible realizations can be found e.g. in \cite{benediktPHD, pulliaPHD, PIMMS-Study, CASBryant:2018}.\
Here we would like to confine the discussion to the design of a slow extraction with the RF-KO scheme.\
In this scheme, the machine's settings are left constant during the extraction and an external electromagnetic wave is used to provide kicks to the beam until single particles slowly reach the unstable part of phase space.\
This scheme is compatible with a machine fully corrected for chromaticity.\

Specifically for this purpose, we will summarize the results of recent investigations, where the nonlinear dynamics near the third-order resonance have been described and illustrate the distinctions between pursuing loss minimization and TRIBs generation.

Our contribution is organized as following: In Section~\ref{sec:Theory} we start by briefly describing the theory of the resonant slow extraction near a third order resonance.\ 
In Section~\ref{sec:SXExamples} we apply the concepts to two exemplary lattices and show the details of the design of the extraction optics.\ 
We continue in Section~\ref{sec:TestBeam} by describing the extraction efficiency and particle extraction rates with the RF-KO method.\ Some specifications and extracted particle rates will be discussed.\ 
We conclude with a summary and discussion of the results.

\section{Theory}
\label{sec:Theory}

\begin{figure}
    \centering
    \includegraphics[width=\linewidth]{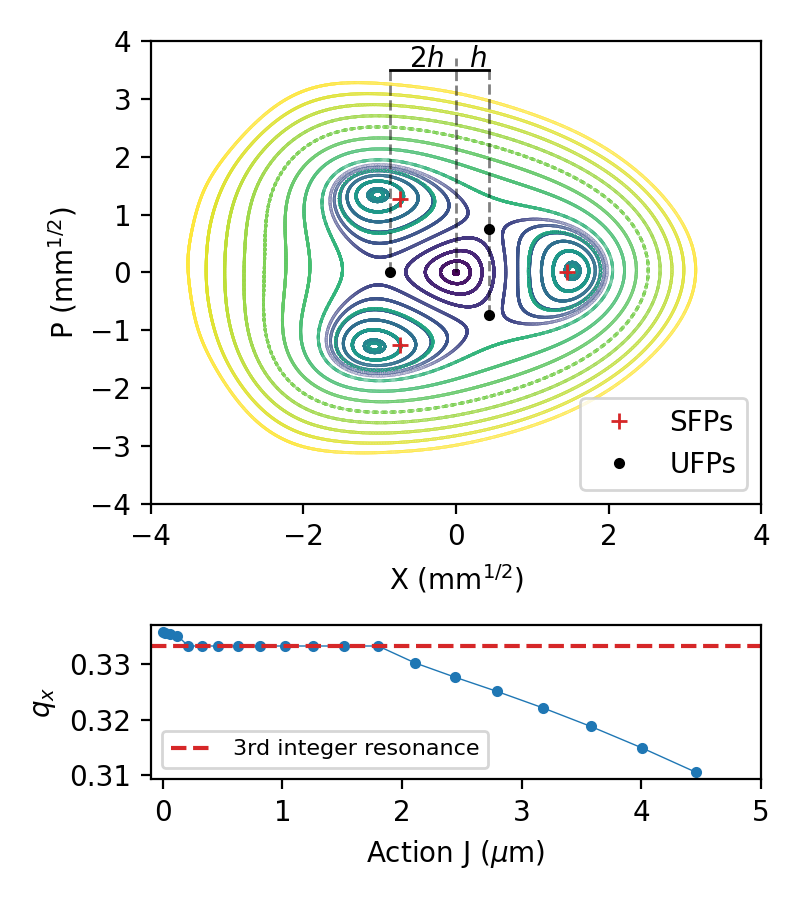}
    \caption{Equipotential lines described by Eq.~(\ref{eq:ResonantHamiltonian}).\ The upper panel shows the result of particle tracking.\ The virtual sextupole strength of the machine was set to $S = $ \SI{30}{\meter}$^{-1/2}$, the tune distance to resonance to $\delta = 2.5 \times 10^{-3}$ and the amplitude dependent tune shift term $\alpha_{xx} = $\SI{-3990}{\meter}$^{-1}$.\ Stable and unstable fixed points (SFP/USP) are also indicated with red crosses and black dots respectively.\
    The lower panel shows the fundamental frequency of motion as a function of the horizontal action.  Horizontal optical functions at observation point are ($\sqrt{\beta_x}$, $\alpha_x$) = (3.975~m$^{1/2}$, 0).}
    \label{fig:phase-space-3h3l-SX}
\end{figure}

\subsection{Effective Hamiltonian near the third order resonance}
For the design of a slow extraction near the third order resonance it is commonly assumed that the horizontal dynamics can be effectively described with the well known Kobayashi Hamiltonian \cite{KobayashiHamiltonian}
\begin{align}
    H &= 6\pi \delta \frac{(X^2 + P^2)}{2} + \frac{S}{4} (3XP^2-X^3),
    \label{eq:KobayashiHamiltonian}
\end{align}
where $(X, P) = (x/\sqrt{\beta_x},\sqrt{\beta_x}p_x + \alpha_x x/\sqrt{\beta_x} )$ are the normalized horizontal coordinates and $\delta = q_x - q_{\text{res}}$ is the tune distance to the nearest 3rd order resonance.\ $\beta_x$, $\alpha_x$ refer to the machine's optical functions from the Courant-Snyder parametrization \cite{COURANT1958}. 
The resonance driving term $S$ can be interpreted as the (normalized) sextupole strength of an effective single virtual sextupole in the lattice.\
The virtual sextupole strength $S$ is given by
\begin{align}
    S &= \big|\sum_n^{N} S_n e^{-i3\mu_n}\big|
    \label{eq:normSextStrength}
\end{align}
with
\begin{align}
    S_n = \frac{1}{2}\int \beta_{x}^{3/2}k_{2}\dd s,
    \label{eq:integratedSextStrength}
\end{align}
where $N$ is the total number of sextupoles in the lattice, \ {$\mu_n = \int_0^{s_n} \dd s / \beta_x$ the betatron phase advance to the $n$-th sextupole and $k_2 = \frac{1}{B\rho}\frac{\partial^2 B_y}{\partial x^2}$ the normalized sextupole field component}.\
Note that Eq.~\ref{eq:KobayashiHamiltonian} is valid in general for on-momentum particles.\
Comprehensive corollaries of the dynamics described by Eq.~(\ref{eq:KobayashiHamiltonian}) and with a more accurate treatment including off-momentum contributions can be found elsewhere (e.g. \cite{pulliaPHD, benediktPHD,PIMMS-Study}) and are the basis of the engineering and optimization of the slow resonant extraction.
For the conventional setup the knobs to control the available dynamic aperture are $S$ and $\delta$.\ The area in phase space where stable motion is possible is given by
\begin{align}
A_{\text{stable}} = 48\sqrt{3}\left(\frac{\pi\delta}{S}\right)^2.
\end{align}
This stable area is a triangle and bounded in the horizontal plane by $X \in [-h_{\text{stable}}, 2h_{\text{stable}}]$ with
\begin{align}
    h_{\text{stable}} = 4\pi\frac{\delta}{S},
    \label{eq:StableArea}
\end{align}
which is the available space for the beam to occupy.\
For the Kobayashi Hamiltonian to be valid two important conditions have to be fulfilled: (a)~The three-turn phase advance should satisfy $6 \pi \delta \ll 1$.\ The distance $\delta$ from the fractional tune $q_x$ to the resonance determines how precise the dynamics can be described with Eq.~(\ref{eq:KobayashiHamiltonian}).\ 
(b) The non-linear dynamics has to be governed by the virtual sextupole $S$.\ Whenever a higher order non-linearity is added, the phase-space structure could be considerably different.\

As previously discussed, in a low emittance lattice the correction of chromaticity has to be performed with high sextupole gradients.\
These can introduce a considerable amplitude dependent tune shift (ADTS) term $\alpha_{xx}$ \cite{SYLeeBook, OctCorr_PRAB21, CornellTRIBS_PRAB23}.\
The amplitude dependent tune shift term reads~\cite{CornellTRIBS_PRAB23}
\begin{align}
    \alpha_{xx} = \alpha_1 + \alpha_2 + \alpha_3,
\label{eq:ADTS_EForest}    
\end{align}
\begin{align}
    \alpha_1 = \frac{1}{2}\left(\cot{\frac{3\Delta}{2}} - \frac{3\Delta}{1-\cos3\Delta} \right) \sum_i^N \sum_j^{j<i} S_i S_j \cos3\psi_{ij},
\end{align}
\begin{align}
    \alpha_2 = \frac{3}{2}\cot{\frac{\mu}{2}}\sum_i^N \sum_j^{j<i} S_i S_j \cos\psi_{ij},
\end{align}
and
\begin{align}
    \alpha_3 = \sum_i^N \sum_j^{j<i} S_i S_j \left[3\sin\psi_{ij} + \sin{3\psi_{ij}}\right],
\end{align}
where $\psi_{ij}$ is the phase advance between the \textit{i}-th and the \textit{j}-th sextupoles, $\mu = 2\pi q_x$, $\Delta = 2\pi \delta$ and $S_{i,j}$ are given by Eq.~(\ref{eq:integratedSextStrength}).

Equation~\ref{eq:ADTS_EForest} cannot recover precisely the ADTS term induced by the set of sextupoles in the machine, nonetheless it gives an insight into how the term is generated.\ 
Therefore we consider the Hamiltonian describing the dynamics near the third order resonance to the lowest order as in \cite{Symon1968beam}
 \begin{align}
     H_r = 6\pi\delta J_x + \frac{S}{\sqrt{2}} J_x^{3/2}\sin{(3\phi_x + 3\mu_S)} + 3\pi\alpha_{xx}J_x^2,
     \label{eq:ResonantHamiltonian}
 \end{align}
with
\begin{align}
    \tan{3\mu_S} &= \sum_n^{N} S_n \sin{3\mu_n}/\sum_n^{N} S_n \cos{3\mu_n}.
\end{align}

Note that the first two terms on the r.h.s of Eq.~(\ref{eq:ResonantHamiltonian}) are simply obtained by introducing action-angle variables $(J_x,\phi_x)$ with $(X, P) = (\sqrt{2J_x}\sin\phi_x, \sqrt{2J_x}\cos\phi_x)$ to Eq.~(\ref{eq:KobayashiHamiltonian}).\
When the sextupole strengths are high the resonance driving terms $S$ and $\alpha_{xx}$ are generated and are not independent. 

The fixed points of Eq.~(\ref{eq:ResonantHamiltonian}) are given by solving for
\begin{align}
    \frac{\partial H_r}{\partial J_x} = 0 =\frac{\partial H_r}{\partial \phi_x},\label{eq:FixedPointEquation}
\end{align}
which then yields
\begin{align}
    J^{1/2}_{x,\pm} = \frac{1}{8\sqrt{2}\pi}\frac{S}{\alpha_{xx}}\left(1 \pm \sqrt{1 - 128\pi^2\frac{\delta \alpha_{xx}}{S^2}}\right),
    \label{eq:ActionFixedPoint}
\end{align}
\begin{align}
    \phi_x = \frac{k\pi}{3} - \mu_S
    \label{eq:PhaseFixedPoint}
\end{align}
for $k=\pm 1, 3$ and $k=\pm 2, 0$.\ A thorough discussion on the properties of the stable and unstable fixed points can be found in \cite{FraserPRAB:2019, CornellTRIBS_PRAB23}.\

With Eq.~(\ref{eq:ActionFixedPoint}) a natural criterion can be established to evaluate if the dynamics of a given lattice can be sufficiently well described by Eq.~(\ref{eq:KobayashiHamiltonian}), namely 
\begin{align}
    \frac{\delta \alpha_{xx}}{S^2} \ll \frac{1}{128\pi^2}.
\end{align}
In this case the Hamiltonian is well approximated by Eq.~(\ref{eq:KobayashiHamiltonian}).\
In Fig.~\ref{fig:phase-space-3h3l-SX} the equipotential lines described by Eq.~(\ref{eq:ResonantHamiltonian}) are illustrated.\ 
These were computed by an element-by-element tracking simulation with \texttt{Xsuite}~\cite{XSuite}.\
Note that the stable and unstable fixed points are also indicated with red crosses and black dots respectively.\ The slight difference from the tracking results come from the fact that higher order non-linear terms cannot be completely neglected \cite{CornellTRIBS_PRAB23}.\

\subsection{Optics design sketch}

For the design of the slow extraction optics, we will pursue the generation of fixed points such that the values given by Eq.~(\ref{eq:ActionFixedPoint}) are real and distinct. Further, the principal considerations for the design include the equilibrium rms beam size $\sigma_x$ and the beam distance to the septum $x_{\text{sep}}$.\
We will set the central bucket to accommodate enough space for a 3$\sigma_x$ beam and attempt to set the opposite stable fixed point as far as possible. The area surrounding the central bucket is determined by the unstable fixed points described by Eq.~(\ref{eq:ActionFixedPoint}).\
These are determined by the achievable set of parameters $(S, \alpha_{xx}, \delta)$.\
In the following, the design philosophy is driven by the pursuit of finding a set of parameters that allow for sufficient dynamic aperture for the beam to be stored while keeping a reasonable distance to the septum.\
This will be achieved by setting the condition 
\begin{align}
    h > 3\sigma_x
\end{align}
with $h = (\beta_x J_{x,-} /2)^{1/2}$ obtained by Eq.~(\ref{eq:ActionFixedPoint}) and $\sigma_x$ referring to the rms beam size of the stored beam.\
This condition arises naturally once it is recognized that the area of the central bucket is delimited by $x\in[-2h, h]$.\ This is illustrated in Fig.~\ref{fig:phase-space-3h3l-SX}.\
The other (stable) fixed point $x_{\text{SFP}} = (2\beta_x J_{x,+})^{1/2}$ will provide a guideline for the distance to the septum $x_{\text{sep}}$.

Finally, if a lattice possesses an unusually high ADTS term or the setup does not allow for the generation of the TRIBs, then a single octupole can be introduced to include a degree of freedom and correct the ADTS term $\alpha_{xx}$.\
The ADTS correction by a thin octupole is given by
\begin{align}
    \alpha_{xx,c} = \frac{K_3 L}{32\pi} \beta_x^2,
    \label{eq:OctCorrection}
\end{align}
where $K_3 = \frac{1}{B\rho}\frac{\partial^3 B_y}{\partial x^3}$ represents the octupole strength.

\section{Slow extraction design examples}
\label{sec:SXExamples}
In this section, we describe the resonant extraction in the proposed low-emittance lattices for the successor of the electron synchrotron DESY II at DESY, Campus Bahrenfeld in Hamburg.\
\ {The proposed machine will primarily operate as a dedicated low-emittance booster for the 4th generation light source, PETRA IV} \cite{agapov2024beamdynamicsperformanceproposed}.\
Design considerations for these lattices can be found elsewhere \cite{chao:ipac2021-tupab023}.\

\subsection{Booster ring lattices}
The first proposed lattice for the booster upgrade has a 6-fold symmetry. 
Design considerations for this lattice are reported in~\cite{chao:ipac2021-tupab023}.\ 
The second one is a lattice with an 8-fold symmetry with an enhanced dynamic aperture \cite{JKeil}.\
Both lattices have a similar performance and are currently considered as successors for the booster ring DESY II.\ 

One important consideration for the new booster design was the existing tunnel geometry.\
The 6-fold symmetric lattice has a quasi-hexagonal geometry, which can only be accommodated for a limited range of circumference values.\
An 8-fold symmetric ring could be more easily installed in the tunnel by scaling the circumference.\ Table~\ref{table:DESY4Specs} lists the relevant parameters of both rings.\
These parameters refer to the operation mode for preparing single bunch beams with up to \SI{1}{\nano\coulomb} for filling the ultra-low-emittance ring PETRA IV and will be considered throughout the study for the design of the slow extraction.\
\begin{table}[h!]
    \caption{Nominal parameters of the proposed DESY IV booster ring lattices and equilibrium parameters at \SI{6}{\giga e \volt}.}
    \label{table:DESY4Specs}
    \centering
\begin{ruledtabular}
\begin{tabular}{llll}
    Parameter & Symbol & 6-fold & 8-fold\\
    Circumference & C & \SI{316.8}{\meter} & \SI{304.8}{\meter} \\
    Tunes$^a$ & $Q_x/Q_y$ & 17.37/12.15 &15.13/9.29\\
    Nat.~chromaticity & $\xi_x/\xi_y$ & -41.8/-13.8& -21.7/-10.4\\
    Mom.~compaction&$\alpha_c$ & $3.17\times 10^{-3}$& $3.66\times 10^{-3}$\\
    Partition number & $\mathcal{J}_x$ & 2.56 & 2.72\\
    Natural emittance & $\epsilon_0$ & \SI{19.1}{\nano \meter\radian} & \SI{19.4}{\nano \meter\radian}\\
    Hor.~damping time & $\tau_x $ & \SI{0.8}{\milli\second} & \SI{0.7}{\milli\second}\\
    Momentum spread & $\Delta p/p$ & $2.64\times 10^{-3}$ & $3.33\times 10^{-3}$\\ 
    \end{tabular}
\end{ruledtabular}\\
\footnotesize{$^a$ Note that for resonant slow extraction the tune is brought even closer to the third order resonance.}\\
\end{table}

For a resonance slow extraction it is necessary to have control of the resonance driving term $S$ described by Eq.~(\ref{eq:normSextStrength}).\ 
The control will be implemented by introducing  two families of harmonic sextupoles in the dispersion-free straights sections.

Given the equilibrium momentum spread values (RMS) listed in Table~\ref{table:DESY4Specs}, the resulting chromatic tune spreads are $(\Delta q_x, \Delta q_y) = (0.11, 0.036)$ for the 6-fold symmetric and $(\Delta q_x, \Delta q_y) = (0.079, 0.035)$ for the 8-fold symmetric lattice.\ 
Such a spread will bring the particles too close to low integer resonances, therefore the chromaticity will be kept fully corrected at the expense of residual $S$ and $\alpha_{xx}$.\

\subsection{6-fold symmetric lattice}

\begin{figure}
    \centering\includegraphics[width=\linewidth]{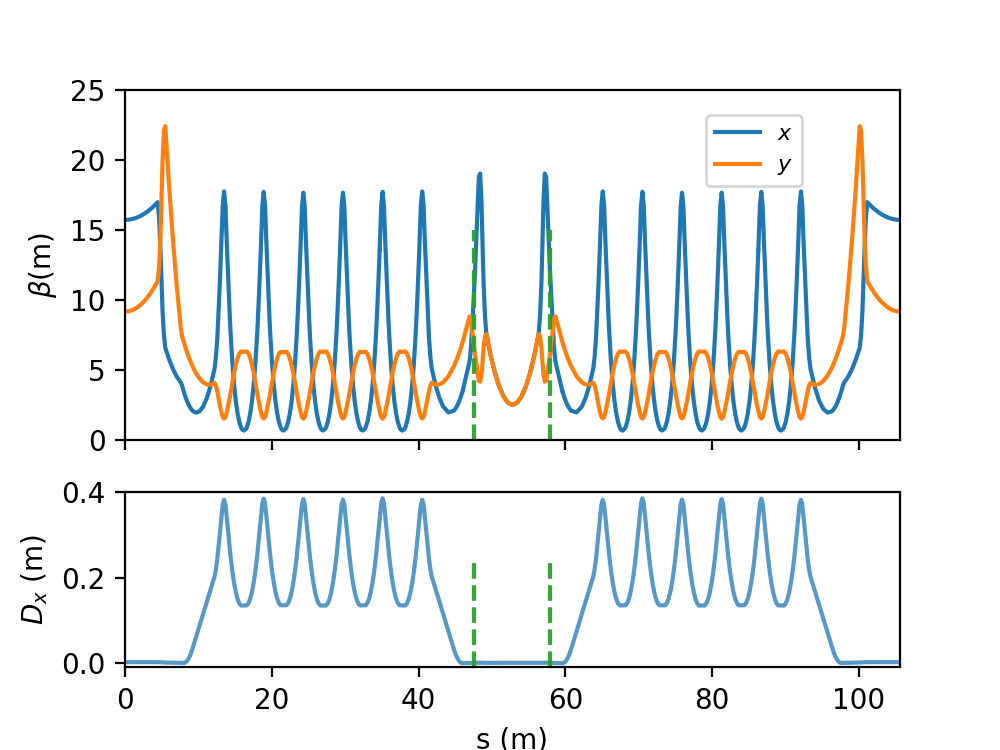}
    \caption{Optical functions of the 6-fold symmetric lattice.\ The dashed green lines indicate the position of two independent harmonic sextupoles.\ The phase advance between them is $\Delta \mu_x \approx \pi/3$.}
    \label{fig:3h3l-Optics}
\end{figure}

The optical functions of one out of three super cells of the full-intensity booster option are shown in Fig.~\ref{fig:3h3l-Optics}.\ 
This lattice is characterized by an alternating array of high-beta and low-beta straight sections, while the arcs remain equal.\ 
The low-beta straights are reserved for the rf cavities.\ 
Injection and extraction devices are supposed to be accommodated at high-beta straights.\
In Fig.~\ref{fig:3h3l-Optics} the position of the harmonic sextupoles is indicated by dashed green lines. The phase advance between them is $\Delta \mu_x = 0.35\bar{3} \pi$.\
Only the available space was used for the integration of the two families of harmonic sextupoles.\

\begin{figure}
    \centering
    \includegraphics[width=\linewidth]{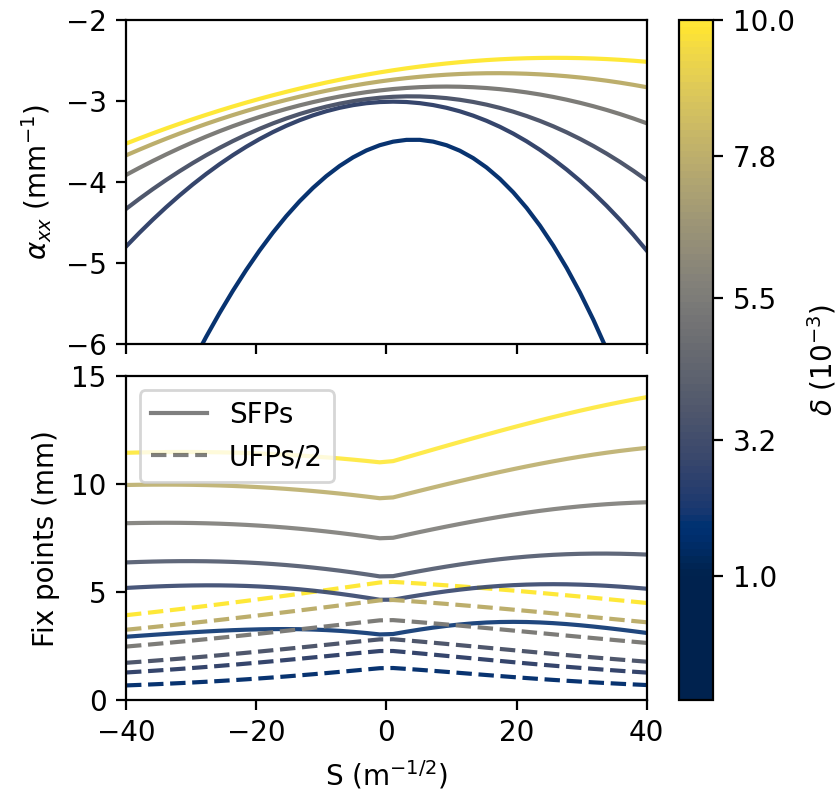}
    \caption{Parameter scan of $(S,\alpha_{xx},\delta)$ for the \ {6-fold symmetric lattice.}\ The upper panel shows the ADTS term as function of the resonance driving term.\ The lower panel shows the corresponding values of the stable and half of the unstable fixed points given by Eq.~(\ref{eq:ActionFixedPoint}) at extraction.}
    \label{fig:AlphaS_scan}
\end{figure}

In order to find a set of optimal parameters $(S, \alpha_{xx}, \delta)$ a systematic scan was performed.\
As discussed in Section~\ref{sec:Theory}, one important aspect for the validity of the dynamics is to be near a resonance.\
Therefore the tune distance to resonance was set to $\delta \ll \frac{1}{6\pi}$.\
The chromaticity was fully corrected to $(\xi_x, \xi_y) = (0, 2)$.\ 

The result of the parameters scan is shown in Fig.~\ref{fig:AlphaS_scan}.\ 
On the top panel the resulting $\alpha_{xx}$ as a function of $S$ is illustrated for $\delta \in [1, 2, 3, 5 , 7.5, 10]\times 10^{-3}$.\
The ADTS term was computed by tracking single particles with different initial actions $J_x$ and fitting the fundamental frequency of motion to a linear function.\
The tracking was performed with \texttt{Xsuite}~\cite{XSuite}.\

The resulting approximate positions of the stable and unstable fixed points are illustrated in the lower panel of Fig.~\ref{fig:AlphaS_scan}, these have been computed with help of Eq.~(\ref{eq:ActionFixedPoint}).\
Observe that the values of Eq.~(\ref{eq:ActionFixedPoint}) are well defined, but the phase of the resonance driving term is not.
The dashed line determines the edge of the central bucket.\
One can establish that for $\delta \geq 2\times10^{-3}$ there is enough space for a 3 rms beam to occupy.\ \ {Note that in the middle of the extraction (high-beta) straight section is} $\sigma_x =$ \SI{560}{\micro\meter}.

To exemplify the phase space structure one set of parameters $(S, \alpha_{xx}, \delta)~=~(30\text{m}^{-1/2}, -3990\text{m}^{-1}, 2.5\times 10^{-3})$ was chosen.\
The ADTS coefficient $\alpha_{xx}$ was calculated with \texttt{PTC-MADX}~\cite{MADX-PTC} as well and show no considerable difference to the value obtained with tracking.\
The resulting phase space portrait is shown in Fig.~\ref{fig:phase-space-3h3l-SX}.\
In Fig.~\ref{fig:phase-space-3h3l-SX} the generation of three TRIBs is clearly visible.\ The stable/unstable fixed points (SFPs/UFPs) computed with Eq.~(\ref{eq:ActionFixedPoint}) are shown.\ 
The slight discrepancy between the expected values come from the fact that higher order ADTS non-linear coefficients and resonance driving terms arise \cite{CornellTRIBS_PRAB23}.\

The unstable fixed points for the chosen set of parameters $(S, \alpha_{xx}, \delta) = (30\text{m}^{-1/2}, -3990\text{m}^{-1}, 2.5\times 10^{-3})$ lie at $(x_{\text{USP}}, -x_{\text{USP}}/2, -x_{\text{USP}}/2)$ = (\SI{-3.4}{\milli\meter},\SI{1.7}{\milli\meter},\SI{1.7}{\milli\meter}) (see Eq.~(\ref{eq:ActionFixedPoint})) and are also made visible in Fig.~\ref{fig:phase-space-3h3l-SX} as black dots.

\subsection{8-fold symmetric lattice}

\begin{figure}
\centering\includegraphics[width=\linewidth]{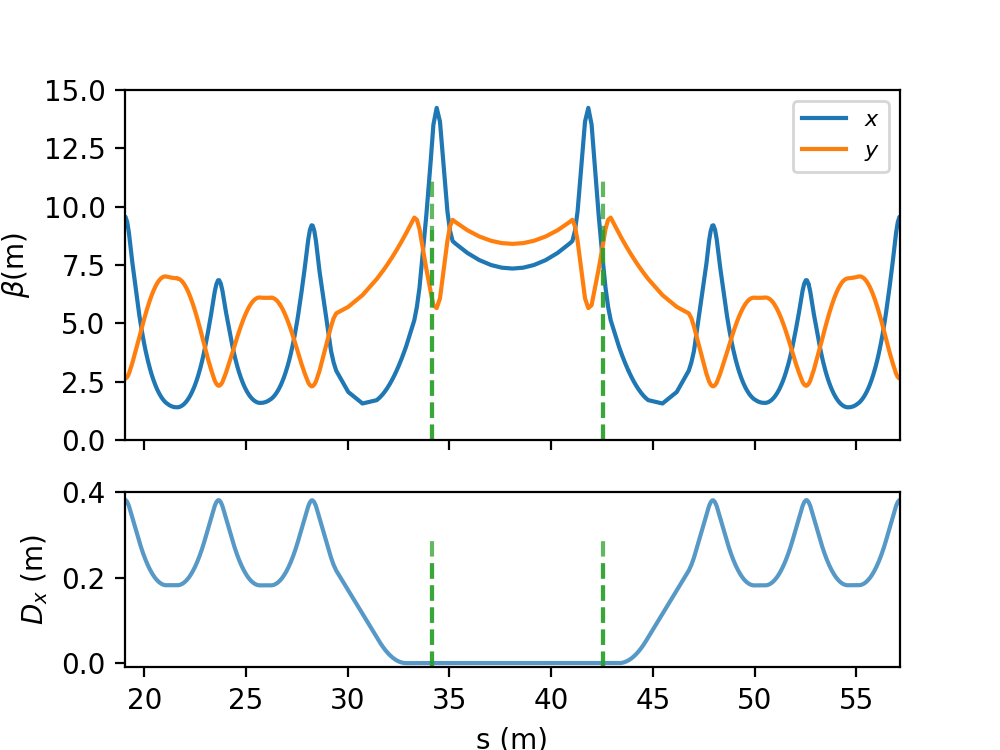}
    \caption{Optical functions of the 8-fold symmetric lattice.\ The dashed green lines indicate the position of two independent harmonic sextupoles.\ The phase advance between them is $\Delta \mu_x \approx \pi/6$.}
    \label{fig:8fold-Optics}
\end{figure}

\begin{figure}
    \centering
    \includegraphics[width=\linewidth]{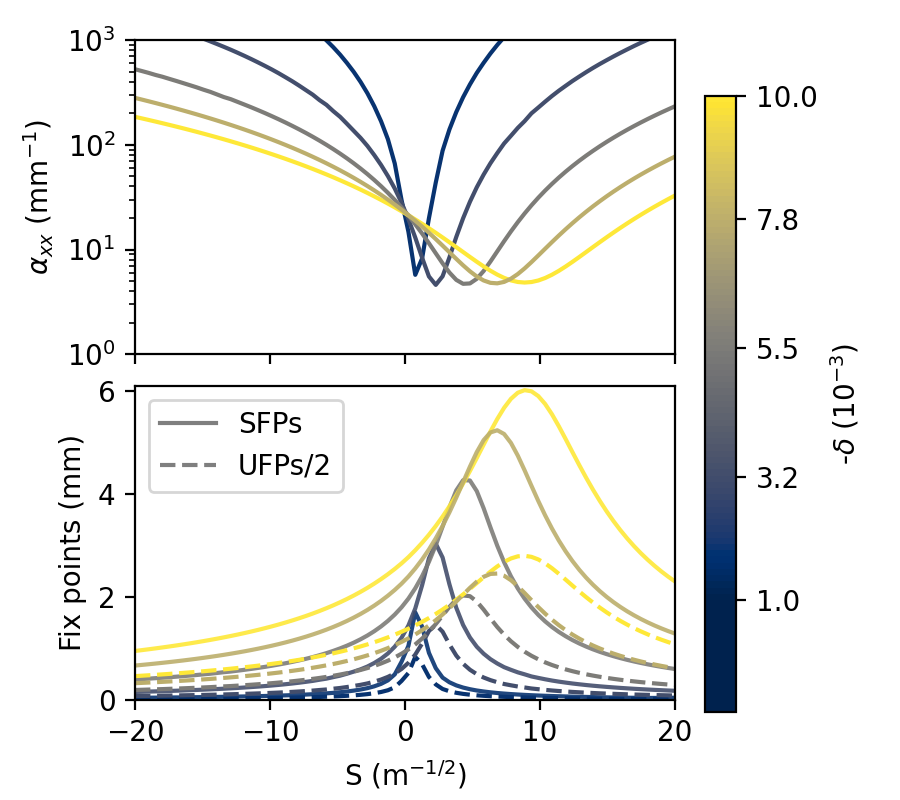}
    \caption{Parameter scan of $(S,\alpha_{xx},-\delta)$ for the 8-fold lattice.\ The upper panel shows the ADTS term as function of the resonance driving term.\ The lower panel shows the corresponding values of the stable and half of the unstable fixed points given by Eq.~(\ref{eq:ActionFixedPoint}).}
    \label{fig:AlphaS_scan_8fold}
\end{figure}

\begin{figure}
    \includegraphics[width=\linewidth]{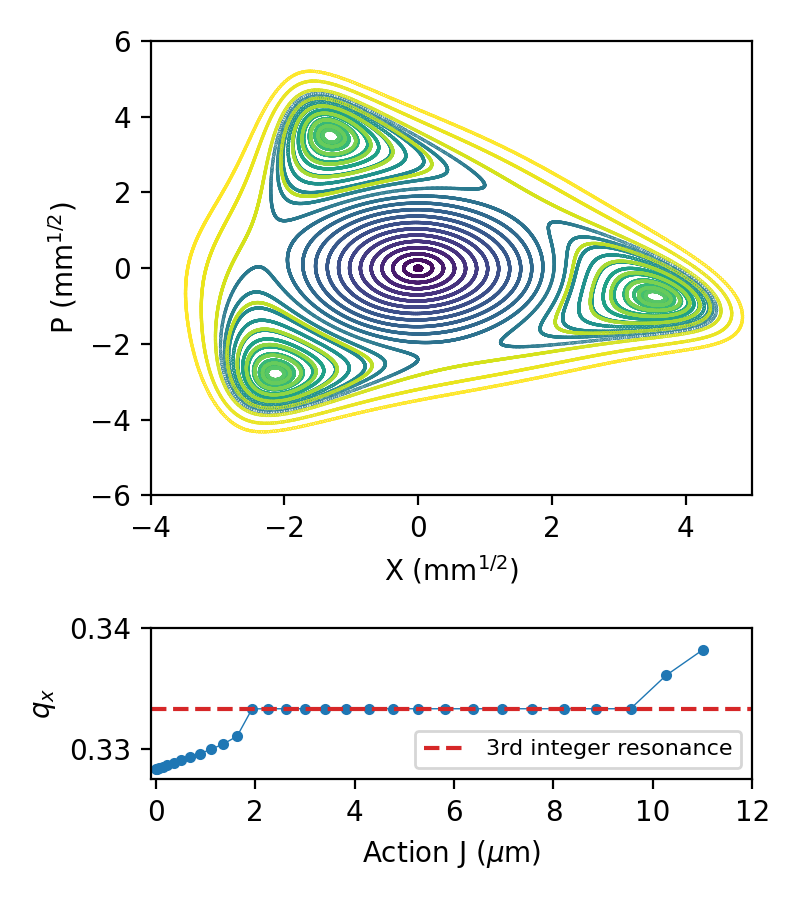}
    \centering
    \caption{Phase space portrait of the 8-fold symmetric lattice at typical slow extraction conditions.\ The upper panel shows the result of tracking.\  The set of parameters was $(S, \alpha_{xx}, \delta) = (5\text{m}^{-1/2}, 2645\text{m}^{-1}, -5\times 10^{-3})$.\ The lower panel shows the fundamental frequency of motion as a function of the horizontal action.\ The generation of TRIBs was only possible with the introduction of a harmonic octupole. Horizontal optical functions at observation point are ($\sqrt{\beta_x}$, $\alpha_x$) = (2.680~m$^{1/2}$,~0).}
    \label{fig:phase-space-8fold-SX-corrected}
\end{figure}

The next candidate lattice is an 8-fold symmetric lattice.\
The optical functions of one super period are illustrated in Fig.~\ref{fig:8fold-Optics}.\
The proposed positions of the harmonic sextupoles are indicated with green dashed lines.\ 
For the preparation of the slow extraction optics the horizontal tune was brought near the third order resonance and the chromaticity was corrected to $(\xi_x, \xi_y) = (0, 1)$.\
To set up the extraction optics the same parameter scan to find an optimal set $(S, \alpha_{xx}, \delta)$ was performed and the results are illustrated in Fig.~\ref{fig:AlphaS_scan_8fold}.\
In the upper panel the magnitude of the ADTS term is shown as a function of the resonance driving term $S$ for $-\delta \in [1, 2.5, 5 , 7.5, 10]\times 10^{-3}$.\
One important consideration for the setup of the extraction settings is the sign of the ADTS term.\
In order for the TRIBs to be formed, the tune as a function of amplitude has to run through the third order resonance.\ In other words, in order for Eq.~(\ref{eq:ActionFixedPoint}) to have two real solutions the sign of $\alpha_{xx}$ and $\delta$ have to be opposite.\
Compared to the previous 6-fold symmetric lattice the ADTS term increases up to two orders of magnitude in the range of the scanned $S$ value.\
The results presented in Fig.~\ref{fig:AlphaS_scan_8fold} suggest that in order to have enough space for the 3$\sigma_x$ beam size the distance to the resonance should be $-\delta > 0.005$.\ \ {The RMS beam size is} $\sigma_x = \SI{373}{\micro\meter}$ in the middle of extraction straight.\
To illustrate the results from the parameter scan shown in Fig.~\ref{fig:AlphaS_scan_8fold} the set $(S, \alpha_{xx}, \delta) = (5\text{m}^{-1/2}, 4070\text{m}^{-1}, -5\times 10^{-3})$ was chosen.\

Interestingly, there appears to be a considerable contribution of higher order ADTS terms that allow the trajectories to cross the resonance and thus prevent resonance islands from forming.\ The phase space portrait is presented in Appendix~\ref{sec:PhaseSpaceUncorrected}.\
This feature is desirable for the booster ring, since in such a lattice crossing low order resonances wouldn't affect the performance.\
On the other hand to prepare the extraction optics, we are looking for a setup where TRIBs can be generated.\

When the presence of higher order ADTS terms beyond $\alpha_{xx}$ define the dynamics, there is still the possibility to find fixed points which have algebraic expressions.\
One can naively consider an extension to the Hamiltonian of the form
\begin{align}
    H = H_r + 2\pi \alpha_{xx}^{(2)} J_x^{3}
    \label{eq:ExtendedHami}
\end{align}
where $H_r$ is given by Eq.~(\ref{eq:ResonantHamiltonian}) and $\alpha_{xx}^{(2)}$ would be a second order ADTS term.\
One can then recover the fixed points by solving Eq.~(\ref{eq:FixedPointEquation}) with this Hamiltonian.\
This in fact yields a depressed quartic equation of the form
\begin{align}
    t^{4} + a t^{2} + b t + c = 0
    \label{eq:DepressedQuartic}
\end{align}
with 
\begin{align*}
    t = J_x^{1/2},&& a = \frac{\alpha_{xx}}{\alpha_{xx}^{(2)}}, && b = \frac{1}{4\sqrt{2}\pi}\frac{S}{\alpha^{(2)}_{xx}}, && c = \frac{\delta}{\alpha^{(2)}_{xx}}.
\end{align*}
The solution to Eq.~(\ref{eq:DepressedQuartic}) yields fixed points that can be expressed analytically \cite{quarticEquation} and are presented in Appendix \ref{sec:QuarticSolution}.\
To this end we recovered the second order ADTS term $\alpha^{(2)}_{xx}$ by correcting the first order term with an octupole.\
Then particles were tracked with different initial actions and a quadratic equation was fitted to the fundamental frequency of motion.\
The second order ADTS term is $\alpha_{xx}^{(2)}=20.7\times10^{6}$~\SI{}{\meter}$^{-2}$.\
We found that the expected fixed points cannot be recovered with help of the solutions of Eq.~(\ref{eq:DepressedQuartic}).\
Nonetheless by introducing the harmonic octupole in the lattice and correcting the ADTS term $\alpha_{xx}$ to approx.~2/3 of the original value ($\alpha_{xx} = \SI{2645}{\meter}^{-1}$ ) the generation of TRIBs was obtained.\ 
The optimal value of correction was found heuristically through trial and error iterations.\ 
It was found that higher corrections of the ADTS term also lead to the formation of TRIBs.\

The resulting phase space portrait with a corrected ADTS is shown in Fig.~\ref{fig:phase-space-8fold-SX-corrected}.\
The formation of the TRIBs is evident and sufficient space for a \SI{20}{\nano\meter\radian} electron beam is available.\ The RMS beam size in normalized phase space is $\sqrt{\epsilon_x} = \sqrt{2} \times 10^{-1}$ mm$^{1/2}$.\ 
Note that since Eq.~\ref{eq:ResonantHamiltonian} is not valid, the control of the phase, i.e. the orientation of the TRIBs is not given by Eq.~\ref{eq:PhaseFixedPoint}.\
Nonetheless the application of \textit{a-posteriori} methods such as multi-objective optimization can aid the control of these knobs~\cite{CornellTRIBS_PRAB23}. 

\section{RF-KO extraction}
\label{sec:TestBeam}

\begin{figure*}
\centering
\includegraphics[width=\linewidth]{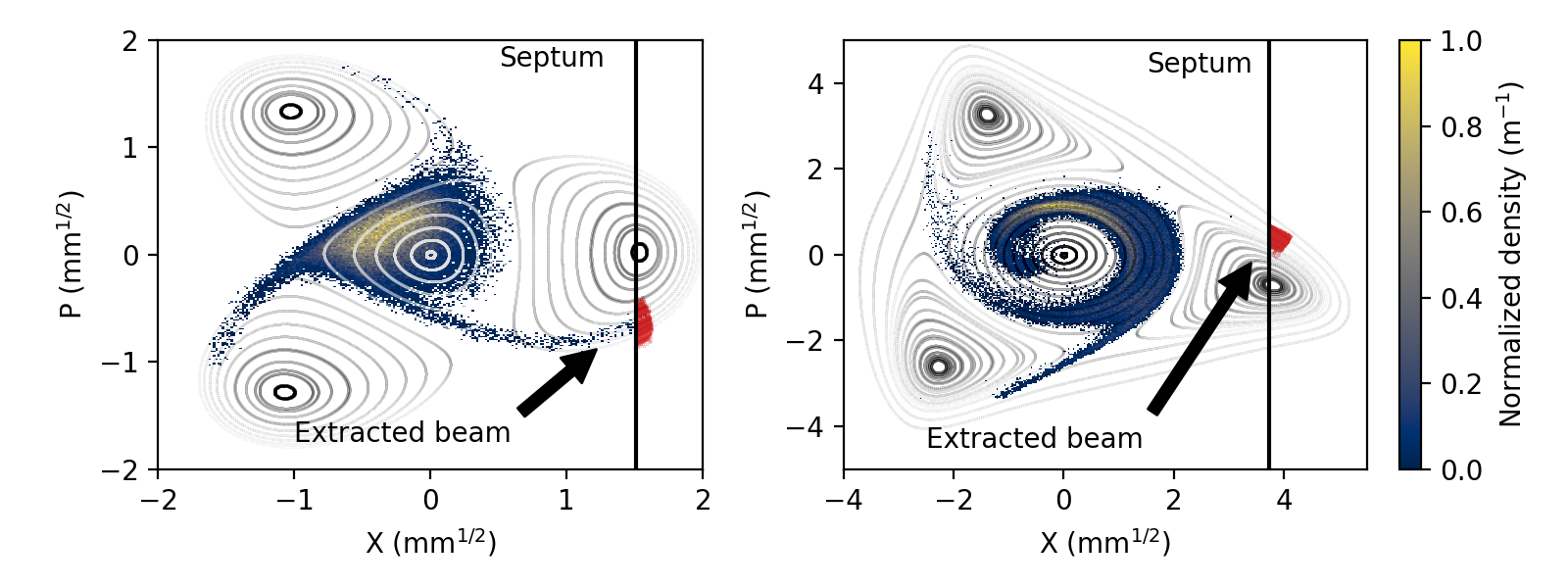}
    \caption{Horizontal phase-space distribution during the slow extraction for the 6-fold (left) and 8-fold (right) symmetric lattices.\ The density of the particle distribution of the circulating beam is shown.\ The extracted particles are the red bulk right of the septum blade (vertical black line).\ The septum was introduced approx.~$x_{\text{sep}} = x_{\text{SFP}}$.\ The equipotential lines are super imposed in the plot.}
    \label{fig:SX-Snapshot}
\end{figure*}

One of the direct applications of the implementation of a slow extraction would be the possibility to continue delivering an extracted electron beam to the HEP test beamlines with the new booster ring.\
DESY operates a test beam user facility, that offers to three independent beamlines electron or positron beam with electable momenta from 1 to 6 GeV/c \cite{D2TestBeam-NIMA2019}.\
These beams are prepared by inserting a thin wire-target (\SI{5}{\micro \meter}) in the way of the circulating electron beam, thus inducing bremsstrahlung that will downstream hit a secondary target to finally produce electron-positron pairs.\

With the upcoming upgrade the test beam facility can potentially profit from a slow extraction implementation.\
The beam users minimum requirements for a potential upgrade include:\
(I) the possibility of a quasi-continuous beam.\
(II) Extraction of particles at the highest energy available is preferred.\
(III) The minimum integrated particle rate should not go below \SI{10}{\kilo\hertz}, which is the integrated rate that is delivered in the current setting.\
(IV) Finally, a flexibility in particle rates is desirable.\ 
For these reasons the slow extraction is perfectly suited for the needs of the test beam user facility.\ 
These considerations also apply to other facilities that can profit from a slowly extracted beam.

Before the slow extraction begins the beam is located in the central bucket after the acceleration ramp.\
During the extraction the initial beam emittance of \SI{20}{\nano\meter} would be slowly blown up with an external excitation wave.\
In addition to conventional RF-KO (with hadron beams), the excitation wave has to provide enough power to counteract the damping of the horizontal betatron motion, such that particles are expelled from the central bucket.\
In this sense, the excitation wave has the function of actively populating the external buckets, but with the additional condition of doing this in a controlled manner and in accordance with the extracted particle rate requested.\

During this process particles would leave the central bucket and either (a) follow a triangular equipotential line surrounding the buckets or (b) become trapped at the edge of an external bucket.\
Due to the influence of synchrotron radiation the horizontal betatron motion will spiral towards a stable fixed point.\ 
For the extraction scheme we are proposing, particles leaving the central bucket would reach the electrostatic septum (ESS) conveniently placed near the stable fixed point of some external bucket and be directed to the extraction channel.\
An illustration of this process is shown in Fig.~\ref{fig:SX-Snapshot}.\
Here a single external sinusoidal wave with the betatron frequency was used to blow up the beam.\ 
The beam distance to the septum was set to approx.~$x_{\text{sep}} = x_{\text{SFP}}$.\
For this simulation, particles are tracked in 6D with synchrotron radiation and quantum excitation effects active.\
The equipotential lines are super-imposed in the picture to illustrate how the particles crossing the separatrix follow the contour of the external buckets.\
Without the septum to stop the motion, the particles would populate the buckets, thanks to synchrotron radiation.\
Note that this exemplary case is meant to illustrate the dynamics of the extraction.\ 
A single sinusoidal wave oscillating at the betatron frequency is known to be far from the optimal excitation wave for RF-KO~\cite{NodaDualFM,NodaRFwithAFM,PNiedermayer2024}.\

\subsection{Extracted beam and particle rates}
\label{sec:ExtractedBeamAndRates}
\begin{table}[ht]
    \caption{Percentage loss due to blade thickness at septum start. A localized aperture restriction at the beginning of the straight section for different thicknesses of the electrostatic septum (ESS) has been implemented. The thicknesses evaluated here correspond to typical specifications for ESS~\cite{CAS2018-Septa, muto:hb2023-frc1i3}.}
\begin{ruledtabular}
\begin{tabular}{c c c c }
    \centering
    \label{tab:ExtractionEfficiency}
    Septum &  \multicolumn{3}{c}{Percentage of particle loss (\%) }\\
    thickness & 6-fold sym.  & \multicolumn{2}{c}{8-fold sym.}  \\
    (\SI{}{\micro\meter}) & $\delta$ = $2.5\times 10^{-3}$ & $\delta$ = $-5\times 10^{-3}$ & $\delta$ = $-0.01$\\
     100 & 35 & 36 & 8\\
      50 & 19 & 14 & 3\\
      30 & 10 & 8  & 2\\ 
\end{tabular}
\end{ruledtabular}

\end{table}

\begin{figure}
    \centering
    \includegraphics[width=\linewidth]{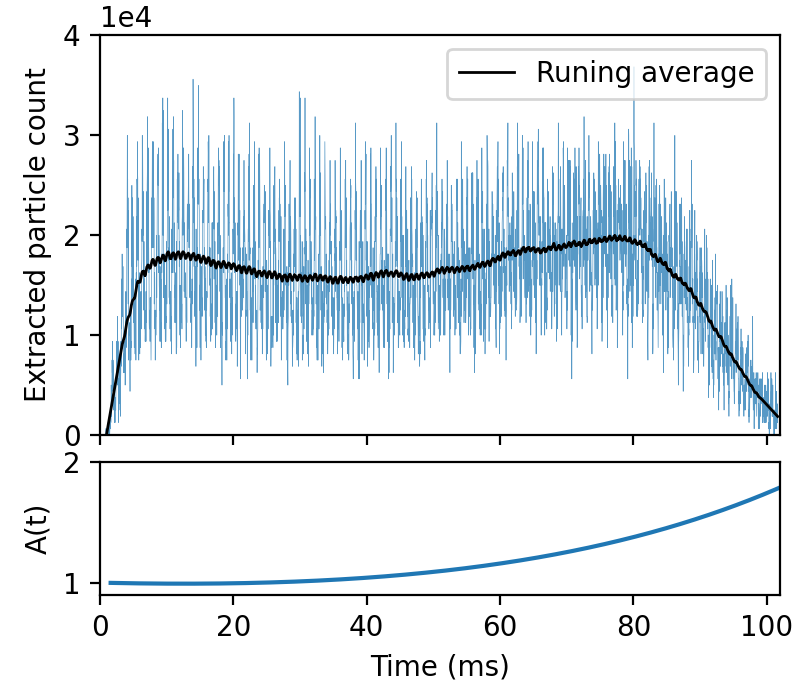}
    \caption{Example of a \SI{100}{\milli \second} slow extraction.\ Extracted particle rates from the 8-fold lattice with $(S, \alpha_{xx}, \delta) = (5\text{m}^{-1/2}, 4070\text{m}^{-1}, -5\times 10^{-3})$ are shown.\ The time bin for counting was set to \SI{25}{\micro\second}.}
    \label{fig:SpillExample}
\end{figure}

\begin{figure}
    \centering
    \includegraphics[width=\linewidth]{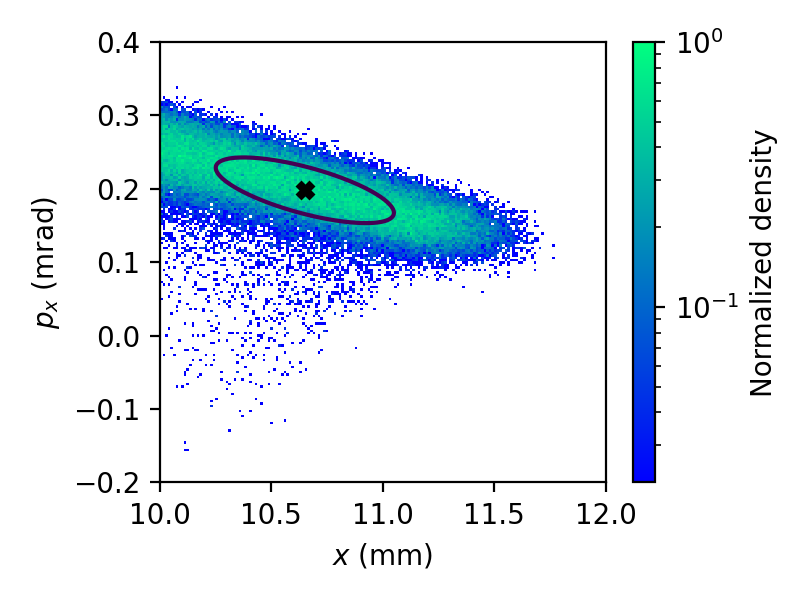}
    \caption{Extracted beam from the 8-fold symmetric lattice with extraction parameters $(S, \alpha_{xx}, \delta) = (5\text{m}^{-1/2}, 4070\text{m}^{-1}, -5\times 10^{-3})$. The ellipse encompasses an emittance $\epsilon_{\text{ext}} =$\SI{13}{\nano\meter\radian} with $\sigma_x = $~\SI{402}{\micro\meter} and $\sigma_{px} =$~\SI{15}{\micro\radian}.}
    \label{fig:ExtractedBeam}
\end{figure}

Extensive particle tracking simulations were performed to evaluate the extracted beam parameters and the extracted particle rate.
First, to evaluate to what extent particles reach the extension of the electrostatic septum (ESS) by a controlled blow-up a single spill of approx.~\SI{100}{\milli \second} was simulated.\
For this simulation the length of the ESS was neglected. Particles were stopped when they reached the extension $x_{\text{sep}} = x_{\text{SFP}}$ at the beginning of the extraction straight section.\ 
An ensemble of $10^{5}$ particles was initialized such that the density distribution is stationary to first order.\
For the 6-fold symmetric lattice (see Fig.~\ref{fig:3h3l-Optics}) $x_{\text{sep}}=$\SI{7}{\milli\meter} and for the 8-fold symmetric lattice $x_{\text{sep}}=$\SI{10}{\milli\meter} (see Fig.~\ref{fig:8fold-Optics}).\
The particles were tracked element by element (thin) with Xsuite~\cite{XSuite}, including synchrotron radiation and quantum excitation effects.\
No misalignment errors were considered.\
An exciter element was introduced to blow up the horizontal emittance.\
The excitation waveform is given by
\begin{align}
    V(t) = A(t) S(t)
\end{align}
with 
\begin{align}
    S(t) &= A_1 \sin{ (2\pi Q_1 f_{\text{rev}} t)} + A_2 \sin{(2\pi Q_2 f_{\text{rev}} t)},\\
    A(t) &= U_1 \exp(-\frac{t}{2\tau_1}) + U_2 \exp( \frac{(t- 1)}{\tau_2}  ),
\end{align}
where the parameters $(Q_1, Q_2, A_1, A_2, U_1, U_2, \tau_1, \tau_2)$ are to be chosen such that the extracted particle rate fulfills the prescribed specifications.\
The parameters $Q_1, Q_2$ define two tune values and control the core diffusion and central bucket escape rate, also known as diffusion and extraction regions in the literature \cite{NodaExtDiffRegion}.\ 
The envelope $A(t)$ of these two sine waves counteract the decaying number of particles getting extracted by adjusting dynamically the excitation strength \cite{FURUKAWA2004}.\ 
This waveform was used for exemplary purposes, since the setup of the parameters is moderately simple.\ 
The introduction of a more complex and effective waveform, such as the ones reported in \cite{PNiedermayer2024}, and its optimization is beyond the scope of this contribution.\

The parameters for the excitation waveform of the 8-fold symmetric lattice were chosen to be $(Q_1, Q_2)~=~(2.3293, 0.328)$, $(A_1, A_2)~=~(2.25, 1.5)$~\SI{}{\micro\radian}, $(U_1, U_2)~=~(1, 10)$ and $(\tau_1, \tau_2)~=~(0.5, 0.25)$.\ 
For the 6-fold symmetric lattice the envelope function was unity and constant along the spill.\
An exemplary spill with the 8-fold symmetric lattice and the amplitude modulation $A(t)$ are shown in Fig.~\ref{fig:SpillExample}.\
The relative extraction rate was multiplied by $N=6.24\times 10^7$, which corresponds to a single bunch current of \SI{10}{\pico \coulomb}.\

For this simulation campaign all particles reached the extension of the ESS. Expected particle losses due to mechanical thickness of the septum blade are listed in Table~\ref{tab:ExtractionEfficiency}.\
An extra set of parameters was included for the 8-fold symmetric lattice with $(S, \alpha_{xx}, \delta) = (10\text{m}^{-1/2}, 2322\text{m}^{-1}, -0.01)$ to show the dependence on $|\delta|$.\
For this setup the ADTS term $\alpha_{xx}$ was diminished with a harmonic octupole correction and again TRIBs were successfully generated.\

Further, an exemplary extracted beam distribution is illustrated in Fig.~\ref{fig:ExtractedBeam}.\
This corresponds to the extracted beam from the 8-fold symmetric lattice with $\delta = 5\times10^{-3}$. One rms beam ellipse has been overlaid to display the emittance, here $\epsilon_{\text{ext}} =$~\SI{13}{\nano\meter\radian}.\
The extracted beam characteristics resemble the "line of charge" expected from the conventional resonant extraction in hadron machines. If compared to the beam characteristics from the example shown in Fig.~\ref{fig:SX-Snapshot}, the emittance is \SI{5.8}{\nano\meter\radian}.\
The six-fold symmetric lattice yields a extracted beam with $\epsilon_{\text{ext}} =$~\SI{1.9}{\nano\meter\radian}.
For the RF-KO method the emittance of the extracted beam can vary depending on the excitation signal~\cite{PNiedermayer2024}. 
Since the endeavour of emittance characterization is closely tied to the investigation of optimal excitation signals, it goes beyond the scope of this paper.

\subsection{Spiral step}

\begin{figure*}
    \centering
    \includegraphics[width=\linewidth]{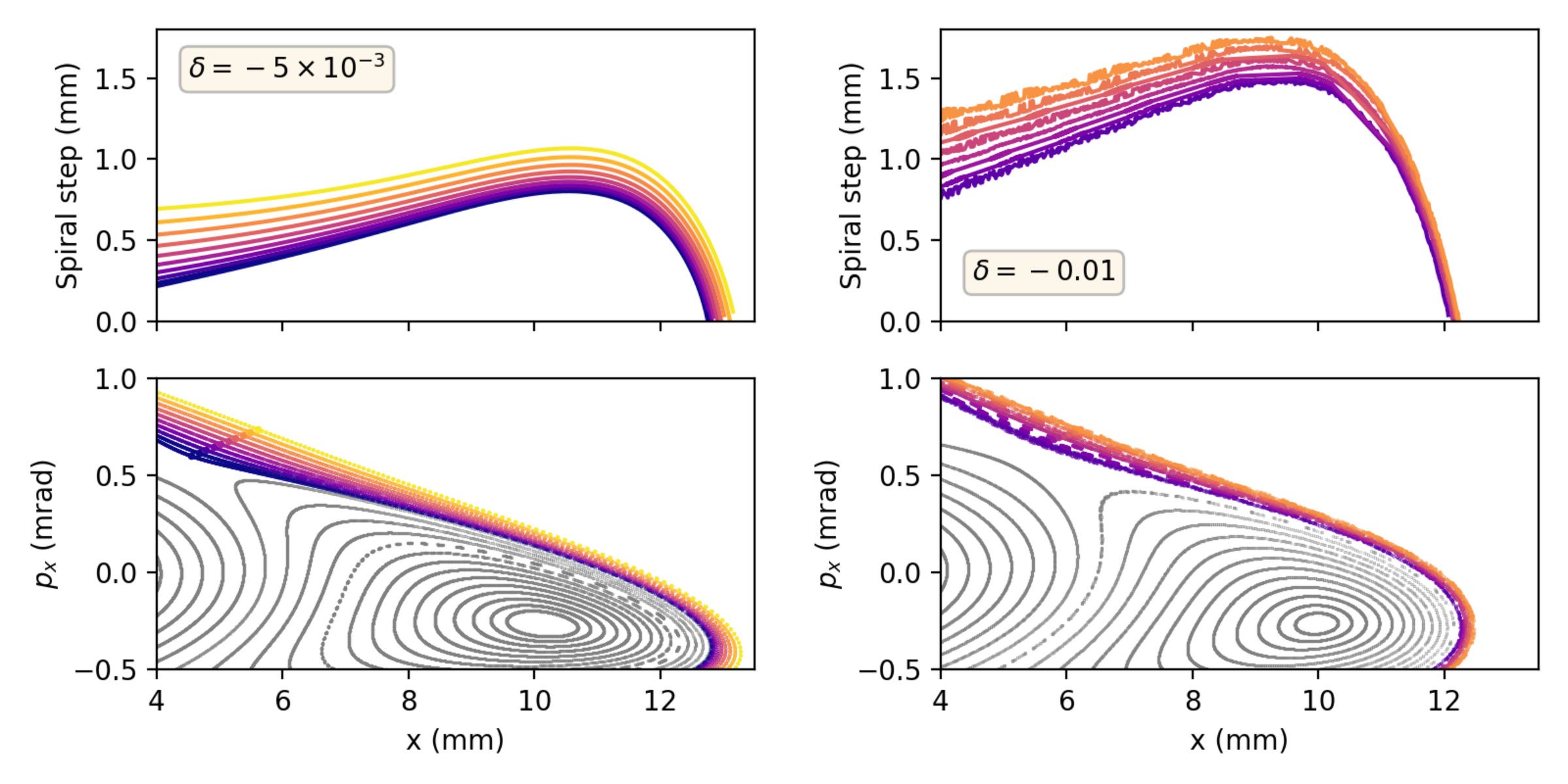}
    \caption{Spiral step for particles just outside the central bucket shown for the 8-fold symmetric lattice on the top panels. On the left panel the extraction settings are $(S, \alpha_{xx}, \delta) = (5\text{m}^{-1/2}, 2645\text{m}^{-1}, -5\times 10^{-3})$. On the right $(S, \alpha_{xx}, \delta) = (10\text{m}^{-1/2}, 2533\text{m}^{-1},  0.01)$. Tracking simulations were performed without synchrotron radiation nor quantum excitation. The lower panels show the corresponding equipotential lines particles follow at the extracted 'arms' shown in Fig.~\ref{fig:SX-Snapshot} (right panel).}
    \label{fig:SpiralStep}
\end{figure*}

The spiral step has been studied using numerical simulations of particles leaving the central bucket (see Fig.~\ref{fig:SX-Snapshot}) for the 8-fold symmetric lattice. This in order to facilitate the discussion of the results presented in subsection~\ref{sec:ExtractedBeamAndRates}.
We define the spiral step in the usual sense, as the (horizontal) amplitude change after three turns~\cite{PIMMS-Study}
\begin{align}
    \Delta x_{\text{spiral}} = x_{n+3} - x_{n}, 
\end{align}
where $n$ represents the turn number. The spiral step for the extracted 'branches' or 'arms' as illustrated in Fig.~\ref{fig:SX-Snapshot} (right panel) is shown in Fig.~\ref{fig:SpiralStep}.\
In Fig.~\ref{fig:SpiralStep}particles are tracked without synchrotron radiation or quantum excitation. Two extraction optics are illustrated. Each curve corresponds to an equipotential line in phase-space, these are depicted in their corresponding lower panels.  Notice that the spiral step increases with $|\delta|$. In the case, where $\delta=-5\times 10^{-3}$ the spiral step still increases beyond the $x_{\text{SFP}}$ suggesting that there is some potential for improvement of the distance from the beam to the septum blade. For the case $\delta=-0.01$ the septum blade at $x_{\text{sep}} = x_{\text{SFP}}$ appears to be a good choice and is also reflected in the values available in Table~\ref{tab:ExtractionEfficiency}.
The results listed in Table~\ref{tab:ExtractionEfficiency} reveal an expected behaviour.\ 
The particle loss decreases as the septum blade thickness decreases.\ 
This is consistent with conventional slow extraction results. The increase of $|\delta|$, increases the probability that a particle will jump over the septum blade and gets extracted.\

A final remark on the influence of synchrotron radiation: as illustrated in~\cite{GentnerPhD}, the relevance of the damping due to synchrotron radiation should be seriously evaluated when $\tau_x \approx T_t$. Here $T_t$ represents the transit time the horizontal amplitude of a particle requires from when it leaves the central bucket to when it reaches the extension of the septum blade.
For the examples we have layout, we estimate the worst case for $\delta = -5\times 10^{-3}$ to be $T_{T} = $~\SI{0.14}{\milli\second}~$\approx \tau_x / 5$.\
This estimation was made by using the information in Fig.~\ref{fig:SpiralStep} and back-tracking a particle from an initial amplitude of 11.5~mm. This information suggests that the influence of radiation damping will be not negligible for particles populating the lower levels of the equipotential lines, when being pushed out of the central bucket. This effect is therefore more accentuated for the 6-fold symmetric lattice and is the reason why a high percentage of particles will get lost at the septum blade. In this example a low $\delta$ leads to a high $T_t$ in the order of $\tau_x$, which then decreases even further the spiral step and thus will have an overall negative impact on the extraction efficiency.

\subsection{Extraction efficiency}

To give an estimate of the extraction efficiency, the ESS was modelled as a \SI{2}{\meter} long blade in the 8-fold symmetric lattice. The length of the ESS has been derived from following tentative considerations: available space of the straight section to be \SI{5.5}{\meter}, the ESS parameters reported in \cite{CAS2018-Septa, muto:hb2023-frc1i3} and a blade thickness of a (Lambertson) magnetic septum downstream of the straight of \SI{4}{\milli\meter}. Simulations show that 1/3 of losses are along the ESS (not at the beginning), mostly downstream at the end of the device.\ A common cure is to consider tilting the ESS, which is equivalent to the introduction of an angle in the closed orbit. Simulations show that a tilt in the orbit of \SI{-250}{\micro\radian} is enough to localize all losses at the beginning of the ESS.
With this information, we conclude that the extraction efficiency for the 8-fold symmetric lattice can be read from Table~\ref{tab:ExtractionEfficiency}, where it can reach the 98\% level for a septum blade thickness of \SI{30}{\micro\meter}. The extracted beam is not divergent (see Fig.~\ref{fig:ExtractedBeam}), so the losses incurred in its transport are kept to a minimum. Note that the the mean angle of the particle distribution is \SI{200}{\micro\radian}. For the 6-fold symmetric lattice the same considerations apply but are not evaluated due to the fact that the extraction optics assessed are in need of optimization, i.e. the evaluation of optics with larger $\delta$.

\section{Summary and conclusion}

In this contribution we proposed a setup of the slow extraction optics at low emittance booster rings by finding the conditions where TRIBs are generated.\ 
We evaluated the systematical generation of TRIB optics by performing a parameter scan.\
To this end, we have evaluated the validity and applicability of Eq.~(\ref{eq:ResonantHamiltonian}) as used in previous studies \cite{CornellTRIBS_PRAB23, FraserPRAB:2019}.\
For one of the evaluated lattices (6-fold symmetric) the obtained results were in good agreement with theoretical predictions.\ The manipulation of the TRIBs could be achieved without octupoles.\ A parameter scan was performed and the range of achievable settings was presented.\

For another candidate lattice (8-fold symmetric) the dynamics cannot be described satisfactorily with help of Eq.~(\ref{eq:ResonantHamiltonian}).\ 
This is due to the fact that higher order non-linearities beyond the first order ADTS term are present in the system.\ The addition of an octupole and correction of the ADTS term showed that with heuristic methods one can generate the TRIBs.\ This was successful for two sets of parameters.\
In several studies the generation of TRIBs \cite{Ries:IPAC2015-MOPWA021, CornellTRIBS_PRAB23, Goslawski:IPAC2019-THYYPLM2} has been successfully experimentally demonstrated.\ 
In this contribution we report a case where the generation of TRIBs would not be possible without the introduction of an additional octupole.\ 
This leads to the conclusion that the Hamiltonian described by Eq.~(\ref{eq:ResonantHamiltonian}) has a regime of validity which is limited and even if the TRIBs can be generated the fixed points won't follow the results of Eq.~(\ref{eq:ActionFixedPoint}).\ 
The evaluation of the punctual regime of validity of Eq.~(\ref{eq:ResonantHamiltonian}) goes beyond the scope of this study.\
A naive attempt was pursued in order to extent and find analytically the fixed points in phase space.\ This was done by introducing a second order ADTS term as shown in Eq.~(\ref{eq:ExtendedHami}).\ 
Introducing higher order resonant and ADTS terms would lead to a situation where no algebraic expressions for the solution of Eq.~(\ref{eq:FixedPointEquation}) are available.\

Moreover, with the TRIBs in place a simulation of an RF-KO extraction was performed and the results were presented.\ 
The extraction efficiency is strongly dependent on the given septum blade thickness and the set spiral step of the optics.\
The extraction efficiency profits tremendously from a high spiral step, specially since particles leaving the central bucket undergo bounded motion.\
Without the introduction of a septum, particles orbits would converge towards a stable fixed point.\
In the usual resonant slow extraction particles leaving the central bucket follow unbounded motion.\
Since the particle dynamics is no different than the conventional extraction, but for particles following the equipotential contours at the edge of the TRIBs instead of a separatrix arm, the conventional wisdom for optimization of the extraction can be applied.\
For instance, the septum blade at the 6-fold symmetric lattice should be positioned on the left, implying that the optimal extraction channel has to be placed inside the synchrotron.\ 
The 8-fold symmetric lattice would profit from a rotation of the phase space in counter clockwise direction, such that the extracted particles meet the ESS with a higher positive angle.\

\ {Furthermore, the simulation shows that the extraction efficiency  can reach values similar to a slow extraction where no TRIBs are generated.}\ \ {Traditional extraction without TRIBs can reach efficiencies at the level of }$>$95\% (see e.g.~\cite{SPSExtractionEfficiency,JPARC_ExtractionEfficiency}). 
\ {In conclusion, the proposed new extraction scheme  can be applied to any low-emittance machine where the generation of TRIBs is possible.}\
\ {The usual resonant slow extraction (without TRIBs) can be implemented whenever the natural chromaticity is low enough to allow operation without its correction.}\
The proposed scheme could be a good option to deliver slowly extracted beam to the test beam users at DESY.\
Moreover, the evaluation of further techniques to increase the extraction efficiency such as other excitation waveforms \cite{PNiedermayer2024} or the introduction of crystal collimators \cite{CrystalColls_PRAB_2023} should be considered.\
Particle loss over the ring was observed to be in the level of 0.1\%.\ It has to be evaluated if a machine protection scheme could localize the losses in a designated collimation region.\
Some technical considerations have to be addressed before the slow extraction can be realized.\ For instance the evaluation of misalignment errors have to be included in the analysis.\ 
The parameters presented can be used to estimate the specifications of the electrostatic and magnetic septa.\
Since the straight sections of these lattices are relatively short, the technical realization might become challenging, especially for the 8-fold symmetric lattice.\
Further investigations have to be performed to estimate the requirements of the power amplifier for the excitation signal, feedback system parameters, the optimal place of the harmonic octupole in case the 8-fold lattice is installed and activation safety hazards of the extraction devices.   

\begin{acknowledgments}
We want to take the opportunity to warmly thank S.~Antipov for carefully proofreading the manuscript and Y.-C.~Chae for fruitful discussions.\
Our gratitude is also extended to the PETRA IV beam physics group for their support, especially to J.~Keil for providing the booster lattices.\ 
Finally we also kindly thank A. Herkert, D. Kim and S. Ackermann for providing information on the test beam facility at DESY.
This research was supported in part through the Maxwell computational resources operated at Deutsches Elektronen-Synchrotron DESY, Hamburg, Germany.
\end{acknowledgments}

\bibliography{main}
\appendix
\section{Phase space portrait of 8-fold symmetric lattice without octupole correction}
\label{sec:PhaseSpaceUncorrected}
\begin{figure}[h!]
    \includegraphics[width=\linewidth]{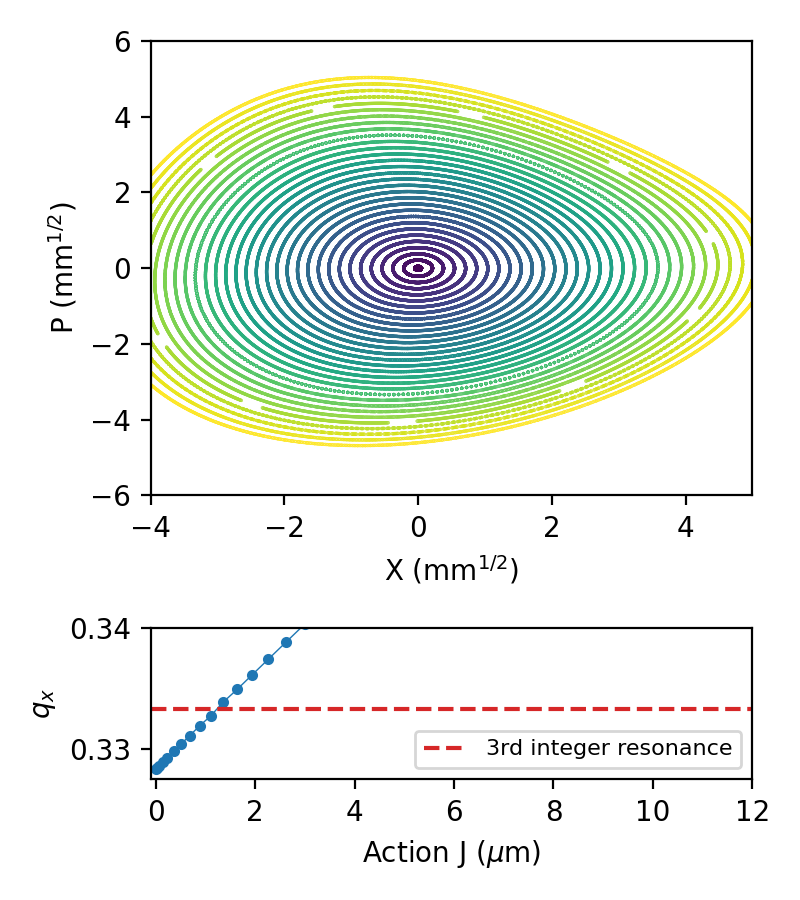}
    \centering
    \caption{Phase space portrait of the 8-fold symmetric lattice without octupole correction.\ The upper panel shows the result of tracking. The set of parameters was to $(S, \alpha_{xx}, \delta) = (5\text{m}^{-1/2}, 4070\text{m}^{-1}, -5\times 10^{-3})$.\ The lower panel shows the fundamental frequency of motion as a function of the horizontal action.\ No TRIBs are generated.}
    \label{fig:phase-space-8fold-uncorrected}
\end{figure}

The phase space with an uncorrected amplitude detuning term is shown in Fig.\ref{fig:phase-space-8fold-uncorrected}.

\section{Solution to the depressed quartic equation}
\label{sec:QuarticSolution}
The solution of the depressed quartic equation
\begin{align*}
    t^{4} + a t^{2} + b t + c = 0
\end{align*}
yields \cite{quarticEquation}
\begin{align}
    t_{\pm} = \frac{1}{2}\left(\sqrt{2y-a}\pm \sqrt{-2y-a-\frac{2b}{\sqrt{2y-a}}}\right)
\end{align}
with
\begin{align*}
    y = \frac{a}{6}+w-\frac{p}{3w}, && w = \sqrt[3]{-\frac{q}{2} + \sqrt{\frac{q^2}{2}+ \frac{p^3}{27}}}\\
    p = -\frac{a^2}{12}-c, && q=-\frac{a^3}{108} +\frac{ac}{3} - \frac{b^2}{8}.
\end{align*}

\end{document}